\documentstyle[aps]{revtex}

\begin{document}
\draft
\title{Navigation in a small world with local information}
\author{Han Zhu$^{\thanks{%
Present address: Department of Physics, Princeton University, Princeton, NJ
08544, USA}}$ and Zhuang-Xiong Huang}
\address{Department of Physics, Nanjing University, Nanjing 210093, China}
\maketitle

\begin{abstract}
It is commonly known that there exist short paths between vertices in a
network showing the small-world effect. Yet vertices, for example, the
individuals living in society, usually are not able to find the shortest
paths, due to the very serious limit of information. To theoretically study
this issue, here the navigation process of launching messages toward
designated targets is investigated on a variant of the one-dimensional
small-world network (SWN). In the network structure considered, the
probability of a shortcut falling between a pair of nodes is proportional to 
$r^{-\alpha }$, where $r$ is the lattice distance between the nodes. When $%
\alpha =0$, it reduces to the SWN model with random shortcuts. The system
shows the dynamic small-world (SW) effect, which is different from the
well-studied static SW effect. We study the effective network diameter, the
path length as a function of the lattice distance, and the dynamics. They
are controlled by multiple parameters, and we use data collapse to show that
the parameters are correlated. The central finding is that, in the
one-dimensional network studied, the dynamic SW effect exists for $0\leq
\alpha \leq 2$. For each given value of $\alpha $ in this region, the point
that the dynamic SW effect arises is $ML^{\prime }\sim 1$, where $M$ is the
number of useful shortcuts and $L^{\prime }$ is the average reduced
(effective) length of them.
\end{abstract}


\section{Introduction}

Stanley Milgram's famous experiment of launching messages toward a target
through acquaintances\cite{milgram} showed that we are living in a small
world\cite{WS,review1,review2,review3,review4MEJ}. This experiment is a
typical example of the various navigation processes taking place on many
social and natural network systems, which are known as small worlds. About
this experiment, as well as its replications on the larger scale\cite
{largescale}, two issues are of special interest\cite
{review4MEJ,Kleinberg,deMoura,ignorance}: First, the existence of short
paths between apparently distant individuals in highly regular network
systems, and second, as first noticed and studied by Kleinberg\cite
{Kleinberg}, the efficiency of network navigators in finding such short
paths.

The first issue has been extensively studied, especially with the
Watts-and-Strogatz Small-World Network (SWN) model\cite{WS} (see Ref. \cite
{review1,review2,review3,review4MEJ} and references therein for review). If
a small portion of long range links are added to a regular network, we now
know that the network diameter, defined as the average shortest path length
between vertices, will grow as $\ln N$, where $N$ is the system size.
Actually, this logarithmic scaling can be proved for a variety of network
models (for example, see Ref. \cite{diameter1,diameter2}), and has also been
observed in various real-world networks\cite{diameterr1,diameterr2}. In some
networks, the diameter increases even slower than $\ln N$. As far as our
knowledge goes, most of the theoretical and experimental works concern the
first issue\cite{review1,review2,review3,review4MEJ}. However, the existence
of a short path itself does not guarantee that a navigator will be able to
easily locate it. With regards to the second issue, we still lack an equally
complete understanding, although the works of Kleinberg\cite{Kleinberg} and
de Moura {\it et al.}\cite{deMoura}, and the recent experiment by Dodds {\it %
et al.}\cite{largescale} have already revealed the interesting and rich
phenomena underlying. In the following we briefly describe the basic idea
about the second issue, and its relationship with the first one. A more
detailed review of what is currently known is given in Sec. \ref{Sec. 2}.

Milgram's experiment probed the structure of the social network by studying
a typical example of the dynamic navigation process. In order to analyze the
second issue, we have to put more emphasis on the dynamics. We can clearly
see in the recent experiment by Dodds {\it et al.}\cite{largescale} that the
navigation process is an interplay between the network structure and the
individuals' decisions based on their limited information. An individual is
far from knowing the whole network, and is therefore impossible to make an
always right decision when forwarding a message\cite{Kleinberg,ignorance}.
Actually, as the Dodds {\it et al}. experiment\cite{largescale} shows,
respondents channel the message through contacts who are {\it considered} to
be the nearest to the target. Thus the actual path length is very likely to
be larger than the shortest one.

In this article, the navigation process in a small world network is
theoretically studied by considering both the static structure and the
dynamic decision-making process. We develop the idea of Kleinberg and
provide a systematic treatment of the model navigation process. In the
one-dimensional case, the path lengths are obtained. When time is taken into
consideration, the dynamics of the navigation process can also be obtained.
The calculation presented could be generalized to systems of higher
dimensions.

This article is organized as the following: The model navigation process in
a small world network is described in Sec. \ref{Sec. 2}. Then the path
lengths are calculated in Sec. \ref{Sec. 3}, and the navigation process is
investigated from the dynamic angle in Sec. \ref{Sec. 4}. Section \ref{Sec.
5} is the summary with some discussions.

\section{The model of the navigation process}

\label{Sec. 2}

First we give the definition of the model. The SWN model presented in a
simple way two intrinsic characteristics of various natural and social
networks in reality, i.e., a high clustering coefficient and a short
diameter. An adding-type one-dimensional SWN model\cite{WS} can be
constructed as the following (see Fig. 1): We start from a closed ring of $N$
vertices with only nearest neighbor connections. The nodes are numbered
sequentially from $0$ to $N-1$. (For simplicity we suppose that $N$ is a
multiple of $4$.) Thus, the lattice distance between two nodes numbered $i$
and $j$ is 
\begin{equation}
r_{i,j}=N/2-\left| \left| i-j\right| -N/2\right| .  \label{rij}
\end{equation}
Then, with probability $p$ ($0\leq p\leq 1$) each vertex is additionally
linked to one of the other nodes (excluding its original nearest neighbors).
If this other node is selected at random, then we shall create a small-world
network with random shortcuts. Here, following the idea of Kleinberg\cite
{Kleinberg}, we shall also study the case when the shortcuts are added in a
biased manner: With node $i$ being one end of the shortcut, the probability
that node $j$ is selected as the other end is a function of the lattice
distance between them, 
\[
\frac{r_{i,j}^{-\alpha }}{\sum_{j\neq i,i\pm 1}r_{i,j}^{-\alpha }}=\frac 1A%
r_{i,j}^{-\alpha }, 
\]
where $\alpha $ is a positive exponent and $1/A$ is the normalization
factor. Obviously, the probability that node $i$ and node $j$ are connected
is 
\[
1-\left( 1-\frac 1Ar_{i,j}^{-\alpha }\right) ^2. 
\]
Previous studies on this variant of the SWN model cover topics such as the
static properties\cite{rangeswn2,rangeswn3}, random walkers\cite{rangeswn1},
and also the navigation process\cite{Kleinberg}. As is shown in Kleinberg's
work\cite{Kleinberg} and below, such a structure bears significant meaning
to the navigation process, and the proper selection of $\alpha $ could
greatly enhance the efficiency.

The navigation process could be simulated with the so-called greedy
algorithm (Fig. 1). Considering the limitation of knowledge, we suppose that
each node has a small range of eyeshot, $v$, i.e., each vertex only has the
information of those vertices that can be reached within $v$ steps. When $%
v=1 $, for example, vertex $A$ sending a message to vertex $B$ first
forwards the message to one of its nearest neighbors, $A_1$, which has the
least lattice distance from $B$ (we suppose that $A$ has only knowledge of
the position of $A_1$). Then $A_1$ forwards the message to $A_2$, ..., until
the message reaches $B$. Obviously, the expected value of the path length
depends on the whole set of parameters: system size $N$, system
dimensionality $D$, the fraction of shortcuts $p$, the range of eyeshot $v$,
the exponent $\alpha $, and the lattice distance $n$.

The previous theoretical works consider $v =1$, although currently we still
need more information to judge whether this can correctly represent the
realistic situation. In the study on a square lattice with $p=1$\cite
{Kleinberg}, Kleinberg proves a lower bound on the average actual path
length $\overline{\left\langle l_\alpha \right\rangle }$ taken by the greedy
algorithm to find a randomly chosen target. The bound is $\overline{%
\left\langle l_\alpha \right\rangle }\geq cN^{\beta \left( \alpha \right) }$%
, where $c$ is a constant, $N$ is the total number of vertices, and\footnote{%
This appears different from Kleinberg's original expression\cite{Kleinberg},
because we have changed the meaning of $N$ from the linear length of the
two-dimensinoal system to the total number of vertices.} 
\[
\beta \left( \alpha \right) =\left\{ 
\begin{array}{c}
\left( 2-\alpha \right) /6,0\leq \alpha <2 \\ 
\left( \alpha -2\right) /2\left( \alpha -1\right) ,\alpha >2
\end{array}
\right. . 
\]
When the shortcuts are added at random, $\overline{\left\langle l_{\alpha
=0}\right\rangle }\geq cN^{1/3}$. $\alpha =D=2$ is a special point, at which
the lower bound of $\overline{\left\langle l_{\alpha =2}\right\rangle }$
grows as $\left( \ln N\right) ^2$. The most striking conclusion of Kleinberg
might be that the navigation process has the highest efficiency only in some
of the small-world network structures. This result could be generalized to $%
D $-dimensional lattice ($D\geq 1$), with the special value of the exponent%
\footnote{%
In a later study of a one-dimensional version concerning the static
properties\cite{rangeswn3}, P. Sen {\it et al. }further point out that the
system shows regularity with $\alpha >D+1=2$, indicating that $\alpha =D+1$
might be a second special point for any dimensionality $D$. This issue will
be addressed in the following discussions.} $\alpha =D$. In the more recent
work of de Moura {\it et al.}\cite{deMoura}, the authors directly study the
average actual path length itself on a Watts-Strogatz SWN model with $\alpha
=0$ and varying $p$, with some approximations. It is found that when the
number of shortcuts is large, the average actual path length grows with the
system size as $N^{1/D\left( D+1\right) } $, i.e., $N^{1/2}$ for the
one-dimensional case and $N^{1/6}$ for the two-dimensional case. It also
depends on the value of $p$, and $\overline{\left\langle l_{\alpha
=0}\right\rangle }\sim \left( N/p\right) ^{1/2}$ is obtained by de Moura 
{\it et al.} for the one-dimensional case.

Usually the concept, small-world (SW) effect, refers to the significant
decrease of the shortest path length (a static property in a given network
structure) by the introduction of a portion of shortcuts, and in this sense
it can be accepted as the static SW effect. In the following we define the
decrease of the actual path length in the dynamic navigation process as the
dynamic SW effect. Here, in the one-dimensional $v=1$ case, we take into
consideration the variation of both $p$ and $\alpha $, and provide a
systematic study of the dynamic SW effect, and especially its threshold. As
we shall see below, the dynamic SW effect arises when total length (or
effective length) of useful shortcuts is comparable to the size of the
network (or the segment under study).

\section{The path lengths}

\label{Sec. 3}We begin with deducing a series of quantities, $\left\langle
l\left( n\right) \right\rangle $, i.e., the expected actual path length
between two vertices separated by $n$ regular bonds. In a network consisting
of $N$ vertices, for $n=1$, we simply have 
\begin{equation}
\left\langle l_\alpha \left( 1\right) \right\rangle =1.  \label{ditui1}
\end{equation}
For $n=2$: With probability 
\[
W_{2\rightarrow 0}=1-\left( 1-p\frac{2^{-\alpha }}A\right) ^2 
\]
the original vertex is directly connected to the target via a shortcut.
Obviously, it is with this probability that the message is transferred
directly to the target. Then, with probability 
\[
W_{2\rightarrow 1}=1-W_{2\rightarrow 0} 
\]
the message would be forwarded along a regular bond, with the path length $%
1+\left\langle l\left( 2-1\right) \right\rangle $. Thus, 
\begin{equation}
\left\langle l_\alpha \left( 2\right) \right\rangle =W_{2\rightarrow
0}+W_{2\rightarrow 1}\left[ 1+\left\langle l_\alpha \left( 1\right)
\right\rangle \right] .  \label{ditui2}
\end{equation}
Now we continue to study the general case. When the message is held by a
node separated from the target by lattice distance $i$ ($1\leq i\leq N/2$), $%
W_{i\rightarrow j}$ denotes the probability that, in the next step, the
message is forwarded to a node separated from the target by lattice distance 
$j$ ($0\leq j\leq i-1$). Applying this set of probabilities, we obtain for $%
2\leq n\leq N/2$, 
\begin{equation}
\left\langle l_\alpha \left( n\right) \right\rangle =W_{n\rightarrow 0}+%
\mathop{\displaystyle \sum }%
\limits_{i=1}^{n-1}W_{n\rightarrow i}\left[ 1+\left\langle l_\alpha \left(
i\right) \right\rangle \right] .  \label{dituin}
\end{equation}
Now we still have to give the explicit expression of $W_{i\rightarrow j}$.
We suppose the lattice distance between the current message holder and the
final target is $i$, and the next holder is separated from the target by
lattice distance $j$. If $j=0$, 
\begin{equation}
W_{i\rightarrow 0}=1-\left( 1-p\frac{i^{-\alpha }}A\right) ^2.  \label{Wi0}
\end{equation}
If $j>0$, the occurrence of this event, $i\rightarrow j$, requires two
conditions to be both satisfied: (1) The current holder must be linked with
the next indicated holder and (2) the current holder cannot be linked with
the other nodes that are closer to the final target than the next holder.
Thus, 
\begin{eqnarray}
W_{i\rightarrow j} &=&\left[ \prod_{k=i-j+1}^{i+j-1}\left( 1-p\frac{%
k^{-\alpha }}A\right) \right] \left\{ p\frac{\left( i-j\right) ^{-\alpha
}+\left( i+j\right) ^{-\alpha }}A\right.  \nonumber \\
&&+\left( 1-p\frac{\sum_{k=i-j}^{i+j}k^{-\alpha }}A\right)  \nonumber \\
&&\times\left. \left[ 1-\left( 1-p\frac{\left( i-j\right) ^{-\alpha }}A%
\right) \left( 1-p\frac{\left( i+j\right) ^{-\alpha }}A\right) \right]
\right\}  \label{Wij}
\end{eqnarray}
for $0\leq j\leq i-2$, and 
\begin{equation}
W_{i\rightarrow i-1}=1-\sum_{j=0}^{i-2}W_{i\rightarrow j}.  \label{Wii-1}
\end{equation}
The exact expression of $W_{i\rightarrow j}$ will be applied in the next
section where the dynamics of the navigation process is studied. But here it
will create technical difficulty for us to obtain the path lengths, due to
the limit of our computational facilities. Instead, we shall employ some
approximations. We consider $p/A$ a relatively small quantity, and if we
retain only the first order terms, Eq. (\ref{Wij}) can be simplified to, 
\begin{equation}
W_{i\rightarrow j}=2p\frac{\left( i-j\right) ^{-\alpha }+\left( i+j\right)
^{-\alpha }}A.  \label{Wij-sim}
\end{equation}

\subsection{The effective diameter}

\label{Sec. 3.1}With Eqs. (\ref{ditui1}) and (\ref{dituin}) we can obtain
the path lengths $\left\langle l_\alpha \left( n\right) \right\rangle $ as a
function of $n$, given any values of the parameters $N$, $p$, and $\alpha $.
But first we do not differentiate among $n$ and study the average actual
path length, $\overline{\left\langle l_\alpha \right\rangle }%
=\sum_n\left\langle l_\alpha \left( n\right) \right\rangle $, in comparison
with the average shortest path length, $\overline{\left\langle d_\alpha
\right\rangle }$. As $\overline{\left\langle d_\alpha \right\rangle }$ is
commonly known as the system diameter, $\overline{\left\langle l_\alpha
\right\rangle }$ can be referred to as the effective diameter for the
navigation process.

In Sec. II, we have presented the concepts of static SW effect and dynamic
SW effect, corresponding to the shortest path length and the actual path
length respectively. The static SW effect is illustrated in Fig. 2 (a $N=1000
$ network with random shortcuts): the diameter is significantly decreased
after a certain threshold region is reached. With respect to the actual path
length, here the result of the effective diameter from Eqs. (\ref{ditui1})
and (\ref{dituin}) is compared with direct numerical simulations in Fig. 2,
and they are in good agreement. Similarly, we also observe the decrease of
the effective diameter in the dynamic navigation process, namely the dynamic
SW effect. The details of this dynamic SW effect, and especially its
threshold, are discussed below.

A very important conclusion has been reached by Kleinberg\cite{Kleinberg}
and de Moura {\it et al.}\cite{deMoura}, in their studies of the variance of
the effective diameter with the system size. As is mentioned above, the
static SW effect is commonly represented in mathematics by a logarithm
scaling, $\overline{\left\langle d\right\rangle}\sim \ln N$. At the same
time, a regular $D$-dimensional lattice has $\overline{\left\langle
d\right\rangle }=\overline{\left\langle l_{regular}\right\rangle }\sim
N^{1/D}$. Kleinberg and de Moura {\it et al.} have found that the dynamic SW
effect lies somewhere between the well studied static SW effect and pure
regularity, as is reviewed in the previous section.

As is mentioned above, the effective diameter is controlled by multiple
factors, and in the following we shall try to include into our discussion
varying $N$, $p$, and $\alpha $, and find out how they are correlated. Given
a network with an arbitrary number of vertices $N$ and an arbitrary value of 
$p$, we reduce the network to unit length. When $\alpha $ is fixed to be $0$%
, we suppose that the structure of the network would be determined by two
values, (1) the number of shortcuts, 
\begin{equation}
M=pN,  \label{M}
\end{equation}
and (2) the average reduced length of the shortcuts, 
\[
L^{\prime }\equiv L/N=\left( N/4\right) /N=1/4, 
\]
where $L$ is the average length of the shortcuts, and the length of a
shortcut is the lattice distance between its two ends, as defined in Eq. (%
\ref{rij}). Since the second factor is a constant, we can consider $M=pN$ as
the only factor that determines the network structure. The reduced effective
diameter $\overline{\left\langle l_{\alpha =0}\right\rangle }/N$ would be a
function of $pN$. We use Eqs. (\ref{ditui1}) and (\ref{dituin}) to prove
this hypothesis. As is shown in Fig. 3 (a) (obtained from Eqs. (\ref{ditui1}%
) and (\ref{dituin})), the data collapse shows, 
\begin{equation}
\overline{\left\langle l_{\alpha =0}^{\prime }\right\rangle }\equiv 
\overline{\left\langle l_{\alpha =0}\right\rangle }/N=f_{\alpha =0}\left(
pN\right) ,
\end{equation}
where $f_{\alpha =0}\left( x\right) \rightarrow 1/4$ for $x\ll 1$, and for
large values of $x$, 
\[
f_{\alpha =0}\left( x\right) \propto 1/\sqrt{x}. 
\]
It means 
\[
\overline{\left\langle l_{\alpha =0}\right\rangle }\propto \sqrt{N/p}, 
\]
and agrees with the result of de Moura {\it et al}\cite{deMoura}. Here we
can see that $pN\sim 1$ is approximately {\it the point where the dynamic SW
effect arises.} (In the following we shall see that actually this point is $%
ML^{\prime}=pN/4\sim 1$.)

We further generalize this hypothesis to arbitrary values of $\alpha $ ($%
\alpha >0$). In the reduced network of unit length, the structure of the
network would be determined by two factors, (1) the number of the shortcuts, 
$M=pN$, and (2) the average reduced length of the shortcuts, which can be
approximately written as 
\begin{eqnarray}
L^{\prime } &=&\frac LN=\frac 1N\frac{\int_1^{N/2}r\times r^{-\alpha }dr}{%
\int_1^{N/2}r^{-\alpha }dr}  \nonumber \\
&=&\left\{ 
\begin{array}{c}
\frac 1N\frac{\alpha -1}{2\left( \alpha -2\right) }\frac{2^\alpha
N^{2-\alpha }-4}{2^\alpha N^{1-\alpha }-2},\left( \alpha \neq 1\right) \\ 
\left( \frac 12-\frac 1N\right) /\ln \left( N/2\right) ,\left( \alpha
=1\right)
\end{array}
\right. .  \label{L}
\end{eqnarray}
For large values of $N$, we have $L^{\prime }\rightarrow const$ ($0<\alpha
<1 $), $L^{\prime }\sim 1/\ln N$ ($\alpha =1$), $L^{\prime }\sim N^{1-\alpha
}$ ($1<\alpha <2$), and $L^{\prime }\sim N^{-1}$ ($\alpha >2$). Now a
question remains to be answered, how are these two factors related? We
further suppose that, with a fixed value of $\alpha $, the network structure
is determined by the direct product of $M$ and $L^{\prime }$. This
hypothesis is supported by our calculations with Eqs. (\ref{ditui1}) and (%
\ref{dituin}), as is shown in Figs. 3 (a)-3 (d). For each value of $\alpha $%
, the data collapse indicates that 
\begin{equation}
\overline{\left\langle l_\alpha \right\rangle }/N=f_\alpha \left( ML^{\prime
}\right).  \label{F}
\end{equation}
When $ML^{\prime }\ll 1$, we always have $f_\alpha \rightarrow 1/4$. It
means that in this region the network remains to be highly regular.

What is especially interesting is $f_{\alpha =1}\left( x\right) $. This is
the point where the dynamic SW effect coincides with the static SW effect.
In the work\cite{Kleinberg} on the two-dimensional network with $p=1$,
Kleinberg obtained 
\begin{equation}
\overline{\left\langle l\right\rangle }\propto \left( \ln N\right) ^2.
\label{ln2N}
\end{equation}
Here, from Eqs. (\ref{M}) and (\ref{L}) we obtain 
\[
ML^{\prime }\approx pN/\ln N\text{,} 
\]
and Fig. 3 (c) shows that 
\[
\overline{\left\langle l_{\alpha =1}^{\prime }\right\rangle }\propto \ln
\left( ML^{\prime }\right) /\left( ML^{\prime }\right) . 
\]
It means 
\begin{eqnarray}
\overline{\left\langle l_{\alpha =1}\right\rangle } &\propto &\frac{\ln N}p%
\left[ \ln p+\ln N-\ln \left( \ln N\right) \right]  \nonumber \\
&\approx &\frac{\ln N}p\left( \ln p+\ln N\right) ,
\end{eqnarray}
which agrees with Eq. (\ref{ln2N}) for $p=1$.

As is pointed out in the previous section, $\alpha =D+1=2$ might be the
other special point. The above analysis can not apply, since there are
virtually no long range links. We have observed that 
\begin{equation}
\overline{\left\langle l_{\alpha >2}\right\rangle }\propto N.
\end{equation}
It is exactly the property of a completely regular network, and agrees with
the conclusion of P. Sen {\it et al.}\cite{rangeswn3}. This issue will be
revisited in the following subsection, where we show that with $\alpha >2$, $%
\left\langle l_{\alpha >2}\left( n\right) \right\rangle \propto n$, another
sign of regularity. (Interestingly, a similar transformation can be found in
a recent work on the aging effect of network systems\cite{aging}.)

\subsection{Path length as a function of lattice distance}

\label{Sec. 3.2}We could have a better understanding of the navigation
process by studying the whole function of $\left\langle l_\alpha \left(
n\right) \right\rangle $, which may help us clarify the relationship between
path lengths and lattice distances. As is shown in Fig. 4, this relationship
is sensitive to the value of $\alpha $.

(1) When $\alpha =0$, the shortcuts are added randomly. We can clearly
identify two distinct regions. If the target is not very far from the source
node, the path length tends to increase linearly with the lattice distance.
This is because, with a small chance of finding a suitable shortcut, the
message is likely to be forwarded solely along regular bonds. However, the
long range shortcuts will dominate the navigation when the target is located
relatively far from the source node. As a result, for most of the region, $%
\left\langle l_{\alpha =0}\left( n\right) \right\rangle $ is highly
independent of $n$. This could be understood with the following qualitative
consideration: Suppose $A$ (or $A_1$) is separated from $B$ by $N/2$ (or $%
N/2-1$) bonds. If no long range bonds exist, the path length from $A_1$ to $B
$ is smaller than that from $A$ to $B$ by one step. However, with long range
bonds in the network, although $A_1$ is closer to $B$ in lattice distance,
the chance of finding a suitable shortcut is also reduced. Our calculation
further shows that these two opposing factors almost counteract each other
completely, and thus $\left\langle l_{\alpha=0} \left( N/2\right)
\right\rangle \approx \left\langle l_{\alpha=0} \left( N/2-1\right)
\right\rangle \approx \cdots $. Obviously, $\overline{\left\langle l_{\alpha
=0}\right\rangle }$ approximately equals the height of the plateau. In Sec. 
\ref{Sec. 4} we shall see that the height of the plateau is just
proportional to $\sqrt{N/p}$, and it directly leads to $\overline{%
\left\langle l_{\alpha =0}\right\rangle }\propto \sqrt{N/p}$. (2) When $%
\alpha =0.5$, the curve of $\left\langle l_{\alpha =0.5}\left( n\right)
\right\rangle $ is similar to that of $\alpha =0$. There is still a range in
which $\left\langle l_{\alpha =0.5}\left( n\right) \right\rangle \propto n$.
But in the plateau that follows, $\left\langle l_{\alpha =0.5}\left(
n\right) \right\rangle $ grows at a very slow, yet detectable pace with $n$.
(3) When $\alpha =1$, the message holders are able to find suitable
shortcuts even when the target is not far away. At the same time, the
shortcuts are not long enough to form a similar plateau as that observed in
the curve of $\alpha =0$. We can observe the following approximate relation, 
\begin{equation}
\left\langle l_{\alpha =1}\left( n\right) \right\rangle \sim \ln n,
\end{equation}
which means 
\[
\frac{d\left\langle l_{\alpha =1}\left( n\right) \right\rangle }{dn}\sim 
\frac 1n. 
\]
(4) When $\alpha >1$, 
\[
\left\langle l_{\alpha >1}\left( n\right) \right\rangle \sim n^\gamma , 
\]
where the exponent $\gamma \approx 0.73$ for $\alpha =1.5$ (given $N=2\times
10^5$ and $p=0.01$)\footnote{%
The value may be different when $N$ and $p$ change. For example, with $%
\alpha =1.5$ and $N=2\times 10^5$ fixed, $\gamma \approx 0.73$ with $p=0.01$%
, $\gamma \approx 0.55$ with $p=0.1$, and $\gamma \approx 0.48$ with $p=1$.
Actually, as discussed below, the value of $\gamma $ is given by $ML^{\prime
}$.}, and $\gamma $ increases to $1$ for $\alpha >2$ (given any values of $N$
and $p$). Since $\left\langle l_{\alpha >2}\left( n\right) \right\rangle
\sim n$ is a property of regular networks, this once again proves the nature
of regularity in the networks generated with $\alpha >2$. The reason may be
that {\it the expected length of shortcuts is finite}. We can also
approximately predict the value of $\left\langle l_{\alpha >2}\left(
n\right) \right\rangle /n$ in the following way. At each time step, with
probability $p$ the message travels along a shortcut of length 
\[
R=\frac{\sum_{r=2}^\infty r^{-\alpha +1}}{\sum_{r=2}^\infty r^{-\alpha }}, 
\]
and with probability $1-p$ the message is forwarded through a regular bond
of unity length. Thus, 
\begin{equation}
\frac{\left\langle l_{\alpha >2}\left( n\right) \right\rangle }n=\frac 1{%
\left( 1-p\right) +pR}.
\end{equation}
This prediction is confirmed by the results of Fig. 4(b).

If we reduce the network to unity length, and plot the reduced length $%
\left\langle l_\alpha ^{\prime }\left( n\right) \right\rangle \equiv
\left\langle l_\alpha \left( n\right) \right\rangle /N$ against $n^{\prime
}\equiv n/N$, we shall be able to observe that data collapse onto a curve,
which is only controlled by $ML^{\prime }$ for each given value of $\alpha $%
. This means that $ML^{\prime }$ gives not only the effective diameter, but
also the function of $\left\langle l_\alpha \left( n\right) \right\rangle $.

\section{The dynamics of the navigation process}

\label{Sec. 4}When a vertex is sending a message to a target located $n$
bonds away, at each time step the message is forwarded to a nearest neighbor
selected based on limited information. In this section we shall turn to
study the dynamics of the navigation process, by calculating the position of
the message as a function of time.

We suppose node $0$ is sending a message to node $n$ ($0<n\leq N/2$). We use
a series of quantities $P_x\left( t\right) $ to denote the probability that
at time $t$ (measured in discrete units) the message is separated from the
target node by $x$ regular bonds. With\ the range of view $v=1$, at $t=0$,
the message is held by the source node and we have

\[
P_n\left( 0\right) =1;P_{x<n}\left( 0\right) =0. 
\]
At $t=1$, the message is forwarded to one of the nearest neighbors of the
source node, and we obtain 
\[
P_n\left( 1\right) =0. 
\]
The probability that the message is forwarded to the node $n-x$ or $n+x$ ($%
n-1\geq x\geq 0$) can be written as, 
\[
P_{n-1\geq x\geq 0}\left( 1\right) =P_n\left( 0\right) W_{n\rightarrow x}, 
\]
where $W_{n\rightarrow x}$ is the probability of the motion and is defined
in Eqs. (\ref{Wij}) and (\ref{Wii-1}). Generally, at time $t$ ($0<t<n$), 
\[
P_{x>n-t}\left( t\right) =0, 
\]
\[
P_{0<x\leq n-t}\left( t\right) =\sum_{y=x+1}^{n-t+1}P_y\left( t-1\right)
W_{y\rightarrow x}, 
\]
and 
\[
P_0\left( t\right) =P_0\left( t-1\right) +P_1\left( t-1\right)
+\sum_{y=2}^{n-t+1}P_y\left( t-1\right) W_{y\rightarrow 0}. 
\]
Finally, at $t=n$, the message completely reaches the target, and 
\[
P_{x>0}\left( n\right) =0;P_0\left( n\right) =1. 
\]
The whole set of probabilities, $P_0\left( t\right) $, $P_1\left( t\right) $%
, ..., $P_n\left( t\right) $ can be obtained, but in the present study we
only use them to calculate the expected position of the message, $%
\left\langle x\left( t\right) \right\rangle $ as a function of time $t$, 
\begin{equation}
\left\langle x\left( t\right) \right\rangle =\sum_{x=0}^nxP_x\left( t\right)
.  \label{xt}
\end{equation}

In the following we try to find out how $\left\langle x\left( t\right)
\right\rangle $ decreases with $t$, and what controls this function. With
dimensionality $D=1$, and the range of view $v=1$, this function still
depends on four parameters: the exponent $\alpha $, the network size $N$,
the fraction of shortcuts $p$, and the lattice distance $n$. In our study of
the path lengths, we cope with this difficulty by reducing each network of
arbitrary size to unit length, and studying accordingly the reduced path
lengths. Using this method we can clearly identify the factors that
determine the network diameter, i.e., the exponent $\alpha $, and a direct
product of the total number of shortcuts and the average reduced bond
length. A similar analysis can be applied to the investigation of the
dynamics.

First we take the networks with $\alpha =0$ as an example. Similarly, we
reduce the segment of network to unit length, and study the reduced function
of $\left\langle x\left( t\right) \right\rangle /n$. We find that with the
value of $\alpha $ fixed, the function of $\left\langle x\left( t\right)
\right\rangle /n$ is only determined by the product of the following two
factors: (1) The first one is the number of useful shortcuts $M_{eff}$. Not
all shortcuts connected to the segment are useful. Only those that can lead
the message to a node closer to the target shall be considered. For example,
with $\alpha =0$, the number of such useful links can be approximately given
by 
\[
M_{eff}\sim pn\frac nN. 
\]
(2) The second factor is the average value of the reduced effective bond
length $L_{eff}^{\prime }\equiv L_{eff}/n$. The effective length of a bond
equals the distance that it can carry the message closer to the target. For
example, with node $0$ as the source node and node $n$ ($0<n<N/2$)as the
target, the effective length of the bond connecting node $n-i$ ($0<i<n$) and
node $n-j$ ($0\leq j<i$) is $i-j$. At the same time, the effective length of
the bond connecting node $n-i$ and $n+j$ is also $i-j$. With $\alpha =0$, we
have approximately 
\[
L_{eff}^{\prime }=L_{eff}/n\sim const. 
\]
The calculations using Eq. (\ref{xt}) support the hypothesis that, with $%
\alpha $ fixed to be zero, the function of $\left\langle x\left(
t\right)\right\rangle/n$ is solely determined by $M_{eff}L_{eff}^{\prime }$, 
\[
\frac{\left\langle x\left( t\right) \right\rangle }n=X_{\alpha
=0,M_{eff}L_{eff}^{\prime }}\left( \frac tn\right) =X_{\alpha
=0,pn^2/N}\left( \frac tn\right) . 
\]
When $pn^2/N\ll 1$, the network is highly regular and obviously $%
\left\langle x\left( t\right) \right\rangle $ will decrease linearly with $t$%
. As $pn^2/N$ increases beyond $1$, the dynamic SW effect arises and we can
observe a faster decay. In Fig. 5, we can see that the initial exponential
decay of $\left\langle x\left( t\right) \right\rangle $ is followed by a
Gaussian cutoff.

This analysis also helps us to understand better the function of $%
\left\langle l_{\alpha =0}\left( n\right) \right\rangle $, which is studied
in Sec. \ref{Sec. 3.2}. With $\alpha =0$, the curve is divided into a region
of linear growth and a plateau, and we can see that the boundary is just $%
pn^2/N\sim 1$.

With other values of the exponent $\alpha $, we can also obtain conveniently
an approximate expression of $M_{eff}$ and $L_{eff}^{\prime }$. In the
preceding paragraphs we have discussed the case of $\alpha =0$. The other
limit case is $\alpha >2$. Obviously, in this case the expected length of
the additional long range bonds is finite, and the network is virtually a
regular one-dimensional ring. If we plot $\left\langle x\left( t\right)
\right\rangle /n$ as a function of $t/n$, we shall observe a linear decay
with the slope larger than one, followed by a plateau where $\left\langle
x\left( t\right) \right\rangle /n$ is almost zero. This is not difficult to
understand, since the effective bond length is larger than unity.

In the region between $\alpha =0$ and $\alpha >2$, it seems difficult to
give a simple characterization of the function of $\left\langle x\left(
t\right) \right\rangle /n$. In this region, the case of $\alpha =1$ is of
special interest. To obtain the exact forms of $M_{eff}\ $and $%
L_{eff}^{\prime }$ we will have to calculate a number of summations, but
here we can conveniently use the following approximate expressions instead
of the exact ones, 
\[
M_{eff}\sim pn\frac{\int_1^n\frac 1rdr}{\int_1^{N/2}\frac 1rdr}=pn\frac{\ln n%
}{\ln N/2}, 
\]
\[
L_{eff}^{\prime }\equiv \frac{L_{eff}^{\prime }}n\sim \frac 1n\frac{\int_1^nr%
\frac 1rdr}{\int_1^n\frac 1rdr}\approx \frac 1{\ln n}, 
\]
and 
\[
M_{eff}\times L_{eff}^{\prime }\sim \frac{pn}{\ln \left( N/2\right) }. 
\]
These expressions are not exact, but they are already able give satisfactory
data collapse. When the dynamic SW effect arises, a typical function is
shown in Fig. 5 (b).

\section{Summary and discussions}

\label{Sec. 5}To summarize, in this article the navigation process is
investigated on a variant of the one-dimensional small-world network (SWN).
In the network structure considered, the long range links are added in a
biased way, i.e., the probability of a shortcut falling between a pair of
nodes goes as $r^{-\alpha }$, where $r$ is the lattice distance between the
nodes. This structure reduces to a SWN with random shortcuts when $\alpha =0$%
. On this network, messages are passed to designated target nodes through
acquaintances. Each message holder forwards the message to one of its
nearest neighbors selected based on its limited information. The system
presents the dynamic small-world (SW) effect, which is defined as the
decrease of the actual path length in the dynamic navigation process by a
portion of shortcuts. This dynamic SW effect is different from the well
studied static SW effect, which refers to the decrease of the shortest path
length. The topics of the present work cover the effective diameter, the
relationship between the path length and the lattice distance, and the
dynamics.

The properties yielded by our calculations are, at the first glance, too
complex to be described by a single theory, due to the multiple parameters,
including $\alpha $, the fraction of shortcuts $p$, the network size $N$,
etc. We provide a unifying analysis, in which we reduce the whole network or
the segment under investigation to unit length, and then accordingly study
the reduced diameter, path lengths, and dynamics. In this way, we use data
collapse to show that the parameters are correlated. This provides us with a
relatively simple method to describe the different aspects of the dynamic SW
effect. The central finding is that, in the one-dimensional network studied,
the dynamic SW effect exists for $0\leq \alpha \leq 2$. With $\alpha >2$,
the system is dominated by regularity. For each given value of $\alpha $
between $0$ and $2$, the point that the dynamic SW effect arises is $%
ML^{\prime }\sim 1 $. If the average actual path length in the whole network
is considered, then $M$ is the total number of shortcuts and $L^{\prime }$
is the average reduced length of them. If only a segment of the network is
considered, then $M$ is the number of useful shortcuts and $L^{\prime }$ is
the average reduced effective length of them (see Sec. IV for definition).
When $ML^{\prime }\ll1$, the system is virtually regular and the navigation
process keeps to be slow. As $ML^{\prime }$ exceeds the threshold of $1$,
the dynamic SW effect arises. The physical meaning of this threshold is also
clear: since $L^{\prime }$ is obtained by dividing the average length (or
effective length) by the size of the network (or the segment under study),
the threshold of the dynamic SW effect is that the total length (or
effective length) of the useful shortcuts is comparable to the network (or
segment) size.

Presently our understanding of the navigation processes and the dynamic SW
effect is far from complete. Related theoretical works also include those on
the scale-free networks and hierarchical structures\cite
{hie,scalefree,powerlaw}. The task is to search for better theoretical
characterization of the navigation processes, find out how they are
influenced by the static properties of the networks, and design network
structures that enable faster navigation. We hope the study on these
problems shall continue to be fruitful.

\acknowledgements

We thank J.-Y. Zhu for helpful discussions.

\null\vskip0.2cm

\centerline{\bf Figure captions} \vskip1cm

Fig.1. A schematic plot of the navigation process with local information.
The information of a vertex is limited by a finite range of view $v$. If $%
v=\infty $ (without any limit), the message will be sent through path I. If $%
v=1$, however, it will be sent through path II, which is longer.

Fig.2. In a $N=1000$ network with random shortcuts ($\alpha=0$), the
diameter ($v=\infty $, squares) and the effective diameter ($v=1$, circles
(simulation data) and solid curve (analytical result)) are plotted as a
function of $p$.

Fig.3 With different values of $\alpha $, (a)-(d) show the relationship
between the reduced effective diameter $\overline{\left\langle l_\alpha
^{\prime }\right\rangle }\equiv \overline{\left\langle l_\alpha
\right\rangle }/N$, and $ML^{\prime }$ (the expressions of which are taken
from Eqs. (\ref{M}) and (\ref{L})). The effective diameters are obtained
from Eqs. (\ref{ditui1}) and (\ref{dituin}). The data collapse in each
subplot consists of $10$ curves with $N=1000$, $2000$, $4000$, ..., $512000$%
, respectively. On each curve with a specific value of $N$, $p=1$, $1.3^{-1}$%
, $1.3^{-2}$, ..., $1.3^{-61}$. Thus there are $10\times 62$ data points in
each subplot. These data collapses strongly suggest Eq. (\ref{F}), i.e., $%
\overline{\left\langle l_\alpha ^{\prime }\right\rangle }=f_\alpha \left(
ML^{\prime }\right) $. (a) $\alpha =0$: $\overline{\left\langle l_{\alpha
=0}^{\prime }\right\rangle }$ is plotted as a function of $ML^{\prime }=pN/4$%
, and the solid line $y\sim \sqrt{x}$ serves as a guide to the eye; (b) $%
\alpha =0.5$: $\overline{\left\langle l_{\alpha =0.5}^{\prime }\right\rangle 
}$ is plotted as a function of $ML^{\prime }\approx pN/6$, and the solid
line represents $y\sim x^{-0.650}$; (c) $\alpha =1$: $ML^{\prime }\approx
pN/\ln N$. In the plot of $\overline{\left\langle l_{\alpha =1}^{\prime
}\right\rangle }\times ML^{\prime }$ as a function of $\ln(ML^{\prime })$,
the right part of the curve appears as a straight line, and this shows that $%
\overline{\left\langle l_{\alpha =1}^{\prime }\right\rangle }\sim \ln \left(
ML^{\prime }\right) /\left( ML^{\prime }\right) $. (d) $\alpha =1.5$: $%
\overline{\left\langle l_{\alpha =1.5}^{\prime }\right\rangle }$ is plotted
as a function of $ML^{\prime }\approx p\left( \sqrt{N}-\sqrt{2}\right) /%
\sqrt{2}$. The solid line represents $y\sim x^{-0.982}$.

Fig.4. In a $N=200,000$ network with $p=0.01$, the path length $\left\langle
l_\alpha \left( n\right) \right\rangle $ obtained from Eq. (\ref{dituin}) is
plotted as a function of $n$ with different values of $\alpha $. In (a), $%
\alpha =0$, $0.5 $, $1.0$, and $1.5$. In (b), the lines with $\alpha =2.1$, $%
2.4$, and $2.7$ are of slope $0.907$, $0.958$, and $0.972$, respectively.

Fig.5. The relationship between $\left\langle x\left( t\right) \right\rangle
/n$ and time $t$, as obtained from Eq. (\ref{xt}). With $\alpha =0$, $%
M_{eff}L_{eff}^{\prime }\sim pn^2/N$, and two sets of parameters leading to
the same value of $M_{eff}L_{eff}^{\prime }$ are chosen: $n=100$, $N=400$, $%
p=0.5$ (squares) and $n=500$, $N=2000$, $p=0.1$ (circles). With $\alpha =1$, 
$M_{eff}L_{eff}^{\prime }\sim pn/\ln \left( N/2\right) $, similarly we
choose:$n=100$, $N=400$, $p=0.3835$ (upward triangles), and $n=500$, $N=2000$%
, $p=0.1$ (downward triangles).

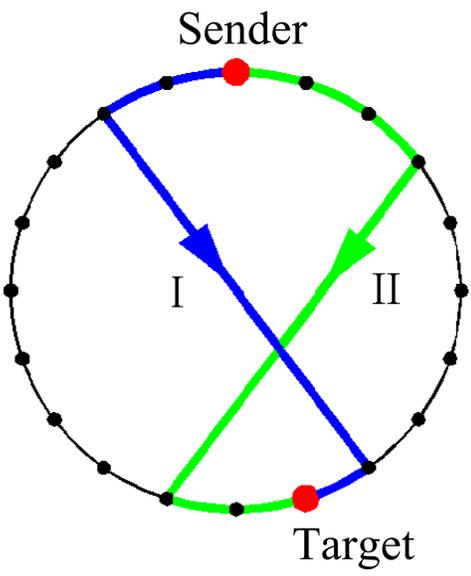
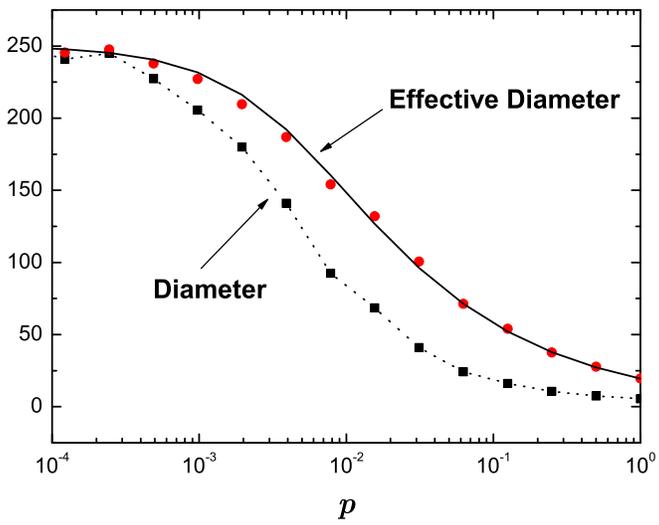
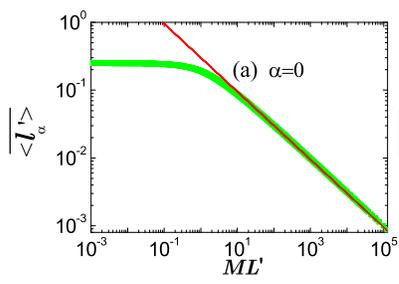
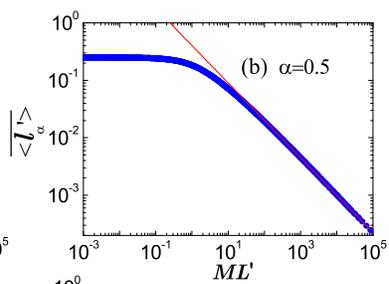
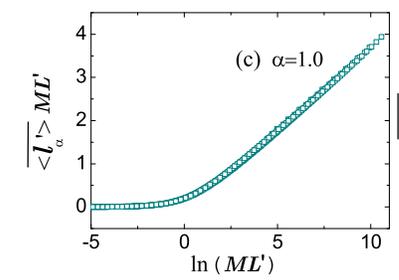
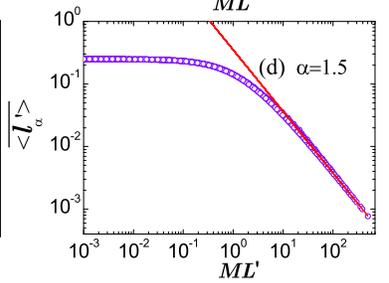
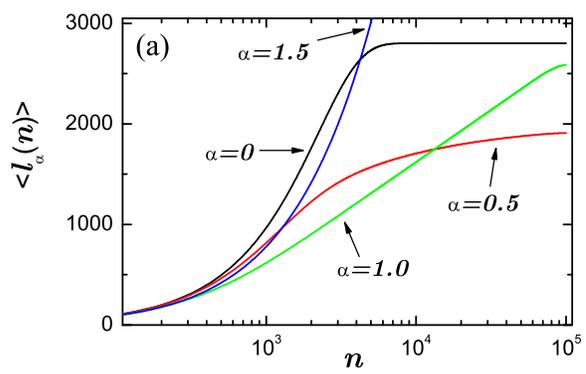
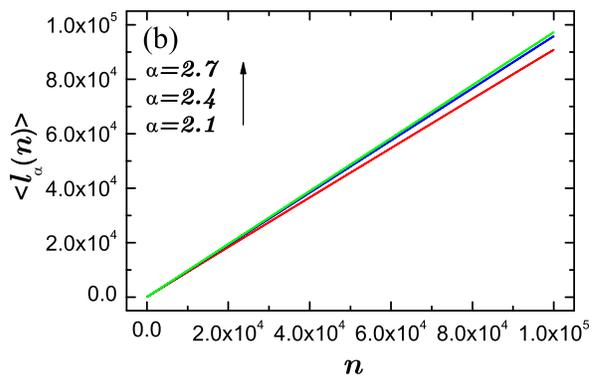
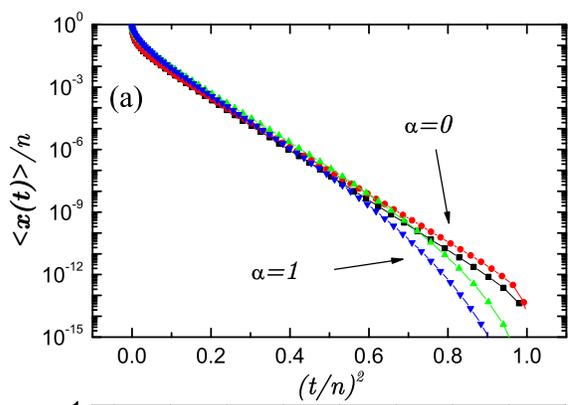
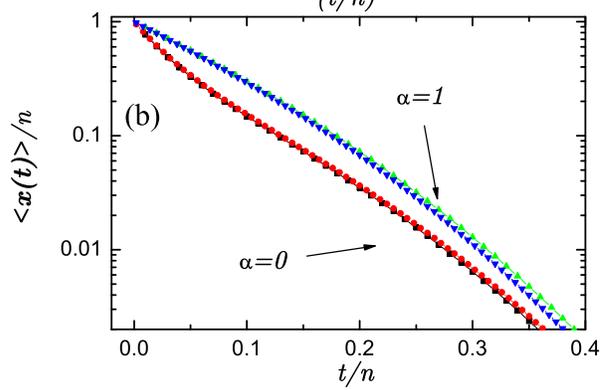
\end{document}